\documentclass[paper]{JHEP3}

\title{A remark on alpha vacua for quantum field theories on de Sitter space}

\author{
Romeo Brunetti and 
Klaus Fredenhagen\\
II. Inst. f. Theoretische Physik  \\
           Universit\"at Hamburg\\
           Luruper Chaussee 149\\
           D-22761 Hamburg, Germany\\
E-mail: \email{romeo.brunetti@desy.de},         
\email{klaus.fredenhagen@desy.de}
}
\author{
Stefan Hollands\\
Inst. f. Theoretische Physik  \\
           Universit\"at G\"ottingen\\
           Friedrich-Hund-Platz 1\\
           D-37077 G\"ottingen, Germany\\
E-mail:
\email{hollands@theorie.physik.uni-goe.de}
}

\abstract {
It is shown that the so-called $\alpha$-vacua which have been proposed as 
candidates for states of free quantum fields on de Sitter space have 
infinitely strong fluctuations for typical observables as the averaged 
renormalized energy momentum tensor.
}

\begin{document}
\section{Introduction}\label{sec:intro}
One of the big problems of quantum field theory on curved spacetime is 
the choice of a state which describes a given physical situation. In particular, 
on a generic (time dependent) spacetime, one cannot single out a state that 
can be viewed as a vacuum state in any natural way. There is, however, strong
evidence that, in the case of free fields, physical states should satisfy 
the so called {\it Hadamard condition} \cite{Brehme-de Wit,Kay-Wald} which 
specifies the singularity structure at coinciding points\footnote{It also 
specifies the singularity structure for lightlike related points that are 
not necessarily close to each other.} but leaves a lot of 
freedom at noncoinciding points. One thus obtains a whole class of 
admissible states, and it was shown, using methods from microlocal analysis 
and concepts from algebraic quantum field theory, that a full fledged 
construction of renormalized perturbative quantum field theory on generic 
backgrounds is possible for such states (see \cite{BF,HW} and references cited there).

On specific spacetimes, one may invoke other criteria to single out states. 
For example, one may ask for states which are invariant 
under the isometries of the spacetime (if any). For instance, on de Sitter space, 
Gibbons and Hawking \cite{Gibbons-Hawking} described an invariant state 
$\omega_0$ which is obtained by a Wick rotation of an invariant state of the corresponding
Euclidean theory. 

Now, for free fields on a spacetime with a PCT symmetry $\Theta$ (like de Sitter 
space where $\Theta$ maps a points $x$ to its antipodal point $x_A$) it is 
possible to obtain from a given representation of the quantum fields a new
representation by applying a so-called ``Bogoliubov automorphism.'' 
For a free hermitian Klein-Gordon field $\varphi$, this is defined by
\begin{displaymath}
\chi_\alpha:        \varphi(x)\to \varphi_{\alpha}(x) \equiv
        \cosh\alpha\,\varphi(x) +\sinh\alpha\,\varphi(\Theta x).
\end{displaymath}   
The field $\varphi_\alpha$ is again a solution to the Klein-Gordon equation, and 
it has the same commutator as $\varphi$, 
  \begin{equation}
    \label{eq:commutator}
    [\varphi(x),\varphi(y)]=i\Delta(x,y)
  \end{equation}
where $\Delta$ is the difference between the retarded and the advanced 
Green's function of the Klein Gordon equation. Namely
\begin{eqnarray*}
    [\varphi_{\alpha}(x),\varphi_{\alpha}(y)]&=&i\cosh^2\alpha\,\Delta(x,y)
    + i\sinh^2\alpha\,\Delta(\Theta x,\Theta y)\\
    &&\qquad +i\sinh\alpha\cosh\alpha\,(\Delta(x,\Theta y)+\Delta(\Theta x,y))\\
    &=& i\Delta(x,y), 
\end{eqnarray*}
where we have used in the last line that 
the PCT transformation $\Theta$ interchanges the retarded and the 
advanced propagator, so that $\Delta(\Theta x,\Theta y)=-\Delta(x,y)$, and where
we have used $\Theta^2=\mathrm{id}$.

The Bogoliubov automorphism $\chi_{\alpha}$ commutes with the action of 
the de Sitter group. Thus, starting from the Gibbons-Hawking 
state $\omega_0$ one 
obtains a 1-parameter family of invariant states\footnote{As shown in~\cite{allen}, 
this family in fact encompasses all de Sitter invariant states.}, the {\it  alpha vacua}
$\omega_\alpha$, which are defined in terms of their $n$-point functions by 
\begin{equation}
  \label{eq:alpha}
  \langle \varphi(x_1)\cdots\varphi(x_n) \rangle_{\omega_\alpha}=
  \langle \varphi_{\alpha}(x_1)\cdots\varphi_{\alpha}(x_n) \rangle_{\omega_0}\ .
\end{equation}
But whereas the Gibbons-Hawking state satisfies the Hadamard condition, 
this is no longer true for the alpha vacua for $\alpha\ne 0$~\cite{allen}.

This can be easily seen in terms of the symmetric part $G_{\alpha}(x,y)$ 
of the 2-point function (the antisymmetric part coincides with the 
commutator and is thus independent of the state). We use the fact that the 
symmetric part $G_0(x,y)$ of the 2-point function of $\omega_0$ is invariant 
under $\Theta$. We thus find
\begin{equation}\label{eq:twopoint}
  G_{\alpha}(x,y)=
  \cosh 2\alpha\,G_0(x,y)+\sinh 2\alpha\,G_0(x,\Theta y)\ .
\end{equation}
Since $G_0$ is singular at coinciding points, it follows 
that $G_\alpha$ is singular also at antipodal points. 
But two Hadamard 2-point functions can differ at most by a smooth function, and because of the hyperbolic prefactors in eq.~(\ref{eq:twopoint}), 
none of the alpha vacua for $\alpha\ne 0$ can be a Hadamard state.

\section{Fluctuations}

We now want to discuss the consequences of the last fact. As a simple case we 
investigate the quantum field which corresponds to $\varphi^2$ in the 
classical theory. The case of energy-momentum tensor, or other field monomials, 
can be treated along similar lines.

It is well known (see, e.g., \cite{BFK}) that in order to define $\varphi^2$ one has to subtract 
a divergent term, for instance the expectation value in the Hadamard state 
$\omega_0$, 
\begin{equation}
  \label{eq:Wick1}
  :\!\varphi^2(x)\!:_{0} \,\, =\lim_{y\to x} \, \varphi(x)\varphi(y)-
                    \langle \varphi(x)\varphi(y) \rangle_{\omega_0} \, .
\end{equation}
The definition depends on the choice of the state used in the subtraction, but the use of a
different Hadamard state leads only to a finite correction. One can even show that 
there is a state independent subtraction which depends only on the 
spacetime metric and its curvature at the point $x$ such that the
expectation values in all Hadamard states are finite. Moreover, averaging 
the field $:\!\varphi^2\!:_{0}$ against a smooth real valued 
test function $f$ of compact support, one obtains a well defined operator 
in the quantum theory, with well defined $n$-point functions in any Hadamard state. 
In particular, this operator has finite fluctuations in such states.

The definition leads, however, to a divergent expectation value 
in an alpha vacuum because of the divergence 
of $G_0$ at coinciding points. One might try another subtraction method 
where instead the expectation value in an alpha vacuum is subtracted.
The quantities $:\!\varphi^2\!:_{\alpha}$ then have well defined 
expectation values in all states in the Fock space built on the same 
alpha vacuum, or, phrased differently, one may insert 
$:\!\varphi^2\!:_{\alpha}$ into an arbitrary $n$-point function of 
an alpha vacuum such as 
\begin{displaymath}
\Big\langle \varphi(x_1)\cdots \varphi(x_n) :\!\varphi^2(y)\!:_{\alpha}
                  \Big\rangle_{\omega_\alpha}
\end{displaymath}
and will obtain a well defined distribution in the variables 
$x_1,\ldots,x_n,y$. But if the averaged field is to be taken seriously as a 
quantum observable we should also investigate its fluctuations. This 
amounts to 2 insertions and will, as we shall see, lead to serious problems.
We may for simplicity look at the fluctuations in the state 
$\omega_{\alpha}$. Since by construction the expectation value in this 
state vanishes, the fluctuations are given in terms of the expectation 
values of the square of the operator. Using the fact that the alpha 
vacuum is a Gaussian state, we have to compute the integral
\begin{equation}
%\int dx dy \, f(x)f(y) \Big\langle 
%:\!\varphi^2(x)\!:_\alpha :\!\varphi^2(y)\!:_\alpha
%\Big\rangle_{\omega_\alpha} \\
%=
\int dx dy \, f(x)f(y)\Big[\langle \varphi(x)\varphi(y)\rangle_{\omega_\alpha} \Big]^2 \ .
\end{equation}
As it stands the expression is not well defined, since a distribution can 
in general not be squared. Since we wish to test the ultraviolet scale
we may look at test functions that are supported in a  
patch that is small compared to the de Sitter scale. Then the leading term in the 
2-point function of $\omega_0$, in normal coordinates around some 
point in the support of $f$, is just the massless 
Minkowski space 2-point function
\begin{displaymath}
  D_+(x,y)=(2\pi)^{-3}\int \frac{d^3{\bf k}}{2|{\bf k}|}e^{ik(x-y)}
\end{displaymath}
where $k=(|{\bf k}|,{\bf k})$. The square of $D_+$ is actually well 
defined which is due to the fact that the convolution of its Fourier 
transform with itself involves only the integration over a compact 
region in momentum space. This is the reason why, in a Hadamard state the 
fluctuations are finite. In an alpha vacuum, however, one obtains also a 
mixed contribution where $D_+(x,y)$ is multiplied with $D_-(x,y)=D_+(y,x)$.
Such a term necessarily diverges; namely
\begin{displaymath}
  \int dxdy \, f(x)f(y)D_+(x,y)D_+(y,x)=\int\frac{d^3{\bf k}}{2|{\bf k}|}
\frac{d^3{\bf p}}{2|{\bf p}|}|\widehat f(k-p)|^2 \ .
\end{displaymath}
Since $f$, being a smooth function that vanishes outside a compact region, 
has a strongly decaying Fourier transform $\widehat f$,
the integration over ${\bf k}$ is essentially restricted to a small 
neighbourhood of ${\bf p}$. It thus amounts essentially to replace 
${|\bf p}|$ by ${|\bf k}|$, and we remain with the linearly diverging integral
\begin{displaymath}
  \mathrm{const}\int\frac{d^3{\bf k}}{2|{\bf k}|^2} \ .
\end{displaymath}

The energy-momentum tensor case can be discussed similarly. The result is a 
fifth power diverging integral. We leave to the reader the easy details.

\section{Conclusions}

We have shown that typical observables have infinitely strong fluctuations in $\alpha$-vacua
for $\alpha\neq 0$.
This holds in particular for the averaged energy-momentum tensor.
The obtained result is valid 
already at the level of free field theory. It is in agreement
with certain results in the existing literature, although it uses
 a much simpler derivation in that it does not rely 
either on perturbation theory \cite{BM,EL} or by coupling the theory 
to gravity \cite{KKLSS}. Therefore, $\alpha$-vacua ($\alpha\neq 0$) are not suited for a 
semiclassical treatment of ``quantum gravity.''
It remains, however,
the possibility to cut-off the mode expansion of the fields at very 
high energy (Planck length), which would
result in a ``better'' short distance behavior of the 
two point function, although breaking de Sitter invariance at such scales. 
This idea seems to be of interest for the purpose of inflation and 
the issue of ``trans-planckian'' physics \cite{D}. 
We shall discuss this additional possibility in a forthcoming contribution.
\section{Acknowledgements}
We are grateful to W. Buchm\"uller who brought to our attention the issue
of alpha-vacua on de Sitter spacetime.


\begin{thebibliography}{[22]}

\bibitem{allen} B.~Allen, \emph{Vacuum States In De Sitter Space}, \prd{32}{1985}{3136}.

\bibitem{Brehme-de Wit} B.S. DeWitt and R.W. Brehme, \emph{Radiation damping in a gravitational field}, \ap{9}{1960}{220}.

\bibitem{BEM} J. de Boer, V. Jejjala and D. Minic, \emph{Alpha-states in de Sitter space}, \hepth{0406217} and references therein.

\bibitem{BF} R. Brunetti and K. Fredenhagen, \emph{Microlocal analysis and interacting quantum field theories: Renormalization on physical backgrounds}, \cmp{208}{2000}{623}.

\bibitem{BFK} R. Brunetti and K. Fredenahgen and M. K\"ohler, \emph{The microlocal spectrum condition and Wick polynomials of free fields on curved spacetimes}, \cmp{180}{1996}{633}.

\bibitem{BM} T. Banks and L. Mannelli, \emph{de Sitter vacua, Renormalization and Locality}, \prd{67}{2003}{065009}.

\bibitem{D} U.H. Danielsson, \emph{Inflation, holography, and the choice of the vacuum in de Sitter 
space}, \jhep{0207}{2002}{040}; \emph{On the consistency of de Sitter vacua}, \jhep{0212}{2002}{025}.

\bibitem{EL} M.B. Einhorn and F. Larsen, \emph{Interacting Quantum Field Theory in
de Sitter Vacua}, \prd{67}{2003}{024001}.

\bibitem{Gibbons-Hawking} G.W. Gibbons and S.W. Hawking, \emph{Cosmologycal event horizon, thermodynamics and particle creation},\prd{15}{1977}{2377}.

\bibitem{HW} S. Hollands and R.M. Wald, \emph{Existence of local covariant time ordered products of quantum fields in curved spacetime}, \cmp{231}{2002}{309}.

\bibitem{Kay-Wald} B. Kay and R.M. Wald, \emph{Theorems on the uniqueness and thermal properties of stationary, nonsingular, quasifree states on spacetimes with abifurcate killing horizon}, \prep{207}{1991}{49}.

\bibitem{KKLSS} N. Kaloper, M. Kleban, A.E. Lawrence, S. Shenker and L. Susskind, \emph{Initial conditions for inflation}, 
\jhep{0211}{2002}{037}.

\end{thebibliography}
\end{document}